\documentclass[prl,twocolumn,showpacs,preprintnumbers,amsmath,amssymb,nofootinbib]{revtex4}

\usepackage[dvips]{graphicx}
\usepackage{subfigure}
\usepackage{rotating}
\usepackage{color}
\usepackage{amsmath,amssymb,hyperref}
\usepackage{hypernat}
\usepackage{ulem}
\DeclareGraphicsExtensions{.eps}

\def\slash#1{#1 \hskip-0.45em /}
\def\beq{\begin{equation}}
\def\eeq{\end{equation}}
\def\bea{\begin{eqnarray}}
\def\eea{\end{eqnarray}}

\def\sectionhead#1{{\it #1}.{\bf ---}}

\newcommand{\newc}{\newcommand}

\newc{\new}[1]{{\color{red}#1}}

\newc{\scale}{0.69}
\newc{\multiplotscale}{0.695}
\newc{\boundaryscale}{0.62}

\newc{\pmea}{{\mathbf p}}
\newc{\pmiss}{\slash{p}}
\newc{\ptmiss}{\slash{p}_T}
\newc{\pya}{p_{Y}}
\newc{\pyb}{p_{Y'}}
\newc{\pna}{p_{N}}
\newc{\pnb}{p_{N'}}
\newc{\pxa}{p_{X}}
\newc{\pxb}{p_{X'}}
\newc{\vlam}{{\bf \Lambda}}
\newc{\lamx}{\lambda_{\tmx^2}}
\newc{\lamn}{\lambda_{\tmn^2}}
\newc{\lamya}{\lambda_{\pya^2}}
\newc{\lamyb}{\lambda_{\pyb^2}}
\newc{\lammya}{\lambda_{\pmiss\pya}}
\newc{\lammyb}{\lambda_{\pmiss\pyb}}
\newc{\lamm}{\lambda_{\pmiss^2}}
\newc{\lamd}{\lambda_{\Delta}}
\newc{\lamP}{\lambda_{P^2}}

\newc{\amm}{\alpha}
\newc{\mya}{\beta}
\newc{\myb}{\beta'}
\newc{\yya}{\epsilon}
\newc{\yyb}{\epsilon'}

\newc{\mx}{m_{X}}
\newc{\mn}{m_{N}}
\newc{\my}{m_{Y}}

\newc{\tm}{\tilde{\bf {m}}}
\newc{\tmmax}{\tilde{\bf {m}}^{\rm max}}
\newc{\truem}{{\bf m}^{\rm true}}
\newc{\tmx}{\tilde{m}_{X}}
\newc{\tmn}{\tilde{m}_{N}}

\newc{\tmnsqmax}{(\tmn^{\rm max})^2}
\newc{\tmxsqmax}{(\tmx^{\rm max})^2}
\newc{\tmnmax}{\tmn^{\rm max}}
\newc{\tmxmax}{\tmx^{\rm max}}
\newc{\tmxmin}{\tmx^{\rm min}}

\newc{\detM}{{\mathcal M}}

\newc{\Wp}{W^+}
\newc{\Wm}{W^-}

\newc{\seR}{\tilde{e}_R}
\newc{\schi}{\tilde{\chi}}

\newc{\mttwo}{m_{T2}}

\newc{\vdot}{\cdot}

\newc{\gam}{\gamma}

\newc{\eg}{e.g.~}
\newc{\ie}{i.e.~}
\newc{\cf}{cf.~}

\newc{\herwig}{\texttt{Herwig++}~}
\newc{\herwigv}{\texttt{Herwig++ v2.5.0}~}

\begin{document}

\preprint{CAVENDISH-HEP-2012-01, LTH-937, DESY 11-233}

\title{A novel technique for measuring masses of a pair of
  semi-invisibly decaying particles}

\author{L.~A.~Harland-Lang$^1$} %\email[]{lucian@hep.phy.cam.ac.uk}
\author{C.~H.~Kom$^{1,2}$} %\email[]{kom@liverpool.ac.uk}
\author{K.~Sakurai$^3$} %\email[]{kazuki.sakurai@desy.de}
\author{W.~J.~Stirling$^1$} %\email[]{wjs2@cam.ac.uk}
\affiliation{$^1$Cavendish Laboratory, J.J. Thomson Avenue, Cambridge
  CB3 0HE, United Kingdom} 
\affiliation{$^2$Department of Mathematical Sciences, University of
  Liverpool, Liverpool L69 7ZL, United Kingdom}
\affiliation{$^3$Deutsches Elektronen-Synchrotron DESY, 22603 Hamburg,
  Germany}

%\date{\today}

\begin{abstract}
  Motivated by evidence for the existence of dark matter, many new
  physics models predict the pair production of new particles,
  followed by the decays into two invisible particles, leading to a
  momentum imbalance in the visible system.  For the cases where all
  four components of the vector sum of the two `missing' momenta are
  measured from the momentum imbalance, we present analytic solutions
  of the final state system in terms of measureable momenta, with the
  mass shell constraints taken into account.  We then introduce new
  variables which allow the masses involved in the new physics
  process, including that of the dark matter particles, to be
  extracted.  These are compared with a selection of variables in the
  literature, and possible applications at lepton and hadron colliders
  are discussed.
\end{abstract}


\maketitle

\sectionhead{Introduction} If new physics (NP) is observed in collider
experiments, the mass of the NP particles involved will be the first
quantities to be measured.  Motivated by the astrophysical evidence of
dark matter, many theories beyond the Standard Model (SM) include a
neutral dark matter (DM) candidate as the lightest of the new
particles.  In many of these models, the stability of the DM against
decays into SM particles is enforced by a new (discrete) symmetry.
Typically such symmetry implies that NP particles are pair produced in
a collider, which subsequently cascade decay into a pair of DM
particles that escape detection.  An example is the minimal
supersymmetric extension of the SM (MSSM) with R-parity.

\begin{figure}[!t]
  \begin{center}
    \scalebox{\scale}{
      \includegraphics{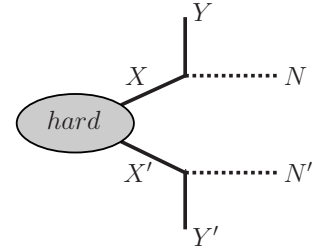}
    }
    \caption{The event topology.  $Y/Y'$ are visible, and their
      4-momenta can be directly observed.  $N/N'$ are dark matter
      candidates; only the vector sum of their 4-momenta could be
      inferred from the momentum imbalance between the initial and
      observed final state particles.}\label{fig:proc}
  \end{center}
\end{figure}

A possible collider process is shown schematically in
Fig.~\ref{fig:proc}.  The NP particle $X/X'$ decays via (a system of)
visible particle(s) $Y/Y'$ into the DM particle $N/N'$.  The momenta
of these particles are denoted $p_{i=X,X',Y,Y',N,N'}$.  If $\pna$ and
$\pnb$ could be measured directly, the {\it true} masses
$\truem\equiv(\mn,\mx)$ for the particles $N/N'$ and $X/X'$ would show
up as delta-function peaks in the invariant mass distributions of
$\pna/\pnb$ and $\pxa/\pxb$ in the limit of zero width and perfect
detector resolution.  In reality, at best the vector sum
$\pmiss=\pna+\pnb$ may be inferred from the 4-momentum imbalance
between the initial state and observed final state particles.  An
observed event is then defined by the 4-momenta set
$\pmea\equiv\{\pya,\pyb,\pmiss\}$.  Although $\truem$ cannot be
measured directly, including mass shell conditions consistent with the
topology in Fig.~\ref{fig:proc} constrains the mass {\it hypothesis}
$\tm\equiv(\tmn,\tmx)$ consistent with $\pmea$ and improves the
determination of $\truem$.  Systematically incorporating these
constraints would hence be beneficial.

In this Letter, we describe a method to determine all possible $\tm$
which takes into account the mass shell constraints when $\pmea$,
 in particular all four components of $\pmiss$, is known, 
 such as at a future linear collider, and in central exclusive production processes at the LHC with tagged forward protons.  For each $\tm$ we obtain analytic solutions for the momenta $p_i$.  Using the
fact that $\truem$ lies within the boundary of $\tm$, we define
boundary variables $\tmmax\equiv(\tmnmax,\tmxmax)$ which develop sharp
edges at $\truem$ without further input.

To illustrate the use of these variables, we will use the example of 
selectron pair production in the MSSM to demonstrate how they complement existing
`standard' mass measurement techniques at future linear colliders,
many of which however do not include information from the mass shell
constraints.  As the edges of $\tmmax$ are independent of the system
centre of mass energy ($\sqrt{s}$), they can be particularly useful at
the Large Hadron Collider (LHC).  We will briefly discuss how our
methods can be used in central exclusive processes, and connections
with `transverse' variables in inelastic processes at the LHC.

%%%%%%%%%%%%%%%%%%%%%%%%%%%%%%%%%%%%%%%%%%%%%%%%%%%%%%%%%%%%%%%%%%%%%%%%

\sectionhead{The calculation method} Given a set of measurable
4-momenta $\pmea$, the 4-momenta of the particles $N,N',X$ and $X'$ in
Fig.~\ref{fig:proc} can be parametrised as
\begin{eqnarray}
  p_{N/N'} &=& \frac{1\mp a}{2}\pmiss \pm\frac{b}{2}\pya \mp\frac{c}{2}\pyb \pm dP\,, \label{eq:nmom}\\
  p_{X/X'} &=& p_{N/N'} + p_{Y/Y'}\,, \label{eq:xmom}
\end{eqnarray}
for dimensionless constants $a,b,c,d$, which includes the missing
momentum constraint $\pmiss=\pna+\pnb$ by construction.  In
Eq.~(\ref{eq:nmom}), the four basis momentum vectors are given by
$\pmea$ and $P$, the latter of which is a {\it space-like} vector
defined by
$P_{\mu}\equiv\epsilon_{\mu\nu\rho\sigma}\pmiss^{\nu}\pya^{\rho}\pyb^{\sigma}$.
As we shall see, the space-like nature of $P$ allows consistent
solutions to be classified using a simple criterion.

The (equal) mass shell constraints are given by
\begin{equation}
  \tmn^2=p^2_{N}=p^2_{N'}\,,\qquad \tmx^2=p^2_{X}=p^2_{X'}\,, \label{eq:mass}
\end{equation}
where $\tm\equiv(\tmn,\tmx)$ are test mass values which need not
coincide with the true masses $\truem\equiv(\mn,\mx)$.  Given $\tm$,
the coefficients $(a,b,c,d)$ can be determined by the four mass shell
conditions.  In fact, using $P\cdot\pmiss=P\cdot\pya=P\cdot\pyb=0$,
three equations linear in $(a,b,c)$ but independent of $d$ can be
obtained by considering the three squared mass differences
$(\pna^2-\pnb^2,\pxa^2-\pna^2,\pxb^2-\pnb^2)$.  Define the Lorentz
invariants
\begin{equation}\label{eq:lambdas}
\vlam\equiv(\lamm,\lammya,\lammyb,\lamya,\lamyb)\equiv(\amm,\mya,\myb,\yya,\yyb)\,, 
\end{equation}
where $\lambda_{p_ip_j} \equiv p_i\cdot p_j/\pya\cdot\pyb$ and
$\lambda_{p_i^2}\equiv p^2_i/\pya\cdot\pyb$.  The solution for
$(a,b,c)$ is then given by
\begin{eqnarray}
  a&=&\frac{1}{\detM}\left[\yya\myb(1+\myb)-\yyb\mya(1+\mya)-\yya\yyb(\mya-\myb)\right]\nonumber \\
  &&+\frac{\lamd}{\detM}\left[\mya(1+\yyb)-\myb(1+\yya)\right]\,, \\
  b&=&\frac{1}{\detM}\left[\yya(1+\myb)(\myb^2-\yyb\amm)+\yyb(1+\mya)(\mya\myb-\amm)\right]\nonumber \\
  &&+\frac{\lamd}{\detM}\left[\amm(1+\yyb)-\myb(\mya+\myb)\right] + \myb\,, \\
  c&=&\frac{1}{\detM}\left[\yyb(1+\mya)(\mya^2-\yya\amm)+\yya(1+\myb)(\mya\myb-\amm)\right]\nonumber \\
  &&+\frac{\lamd}{\detM}\left[\amm(1+\yya)-\mya(\myb+\mya)\right] + \mya\,, 
\end{eqnarray}
where $\lamd\equiv\lamx-\lamn$, and
\begin{equation}
  \detM = 2\mya\myb-\amm(1-\yya\yyb)-\yya\myb^2-\yyb\mya^2
\end{equation}
is the determinant involved when inverting the system of three linear
equations.  Inserting these solutions back into the mass shell
constraints leads to the equation
\begin{eqnarray}\label{eq:quad}
\lamn&=&\frac{c_a}{4\detM}\lamd^2+\frac{c_b}{2\detM}\lamd+\frac{c_c}{4\detM}+d^2\lamP\,.
\end{eqnarray}
This is our main result, from which all variables of interest that we discuss below can be derived.
 The coefficients are given by
\begin{eqnarray}
  c_a &=& \amm(2+\yya+\yyb)-(\mya+\myb)^2 \,,\\
  c_b &=& (\mya+\myb)\left[(\mya+\yya)\myb+(\myb+\yyb)\mya\right]\nonumber \\ 
  &&-\amm[(\mya+\yya)(1+\yyb)+(\myb+\yyb)(1+\yya)]\,, \\
  c_c &=& -\amm^2(1-\yya\yyb)-\left[2\mya\myb+\yya\myb+\yyb\mya\right]^2+4\amm\mya\myb \nonumber \\
  &&+\amm\left[\yyb(1+\yya)(2\mya+\yya)+\yya(1+\yyb)(2\myb+\yyb)\right].~~
\end{eqnarray}
A hypothesis $\tm$ is consistent if the corresponding $\lamd$ and
$\lamn$ lead to $d^2>0$ in Eq.~(\ref{eq:quad}).  In this case a two
fold degenerate solution for $p_i$ with unique $(a,b,c)$ and
$d=\pm\sqrt{d^2}$ is obtained.

We have therefore found a simple criterion to determine the consistency
of $\tm$ with a given $\pmea$, and solve for $p_i$ explicitly in terms
of the Lorentz invariants $\vlam$. More observations on the properties of the solutions can be made.
First, the sign of the energy component of the two solutions (for
$\pm\sqrt{d^2}$) must be the same, since it is always possible to
boost to a frame where the energy component of the space-like vector
$P$ is zero, in which case the two solutions have the same energies.
Second, since the consistent solutions are continuous functions of
$\tmn$ and $\tmx$, the energies of all consistent solutions must have
the same sign.  The energies must then be positive because $\truem$ is
a consistent solution.

It can be shown that $c_a/4\detM<0$ in Eq.~(\ref{eq:quad}).  Since $P$
is space-like, we have $\lamP<0$ and so on the
$(\tmx^2-\tmn^2,\tmn^2)$ plane, the consistent mass region is bounded
from above by Eq.~(\ref{eq:quad}) with $d^2=0$.  Also, it is bounded
from below by $\tmn^2>0$.  This consistent mass region can be
transformed into a corresponding region in the $\tm$ space, which will
be different for each event but will always include $\truem$ in the
absence of detector smearing effects.  A density plot for consistent
mass hypothesis, which in principle includes all kinematic
information, will develop a peaking structure around $\truem$ when a
sufficient number of events are accumulated.

\begin{figure}[!t]
  \begin{center}
    \scalebox{\boundaryscale}{
      \includegraphics{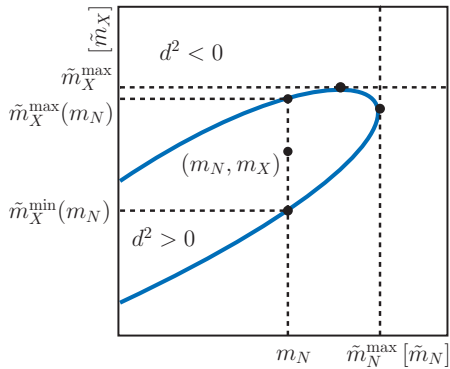}
    }
    \caption{Consistent $(\tmn,\tmx)$ region for a `typical event',
      defined by the 4-momenta $(\pya,\pyb,\pmiss)$.  The region
      $d^2>0$ is consistent. It includes the true mass point
      $(\mn,\mx)$.  $\tilde{m}^{\rm max}_{N,X}$ are the maximum
        $\tmn$/$\tmx$ values, while $\tmx^{\rm min,max}(\mn)$ is the
      minimal/maximal value of $\tmx$ given $\mn$.}\label{fig:typical}
  \end{center}
\end{figure}

Since all solutions $p_i$ consistent with $\pmea$ can now be obtained
for each event, our method provides a departure point for further
analysis of the hard process.  The simple consistent mass boundary
also allows new kinematic variables characterising the mass scales of
the system to be constructed without additional input such as $\mn$.
In particular, the fact that the finite consistent mass region is
characterised by the quadratic curve Eq.~(\ref{eq:quad}) implies that
the maximum consistent values of $\tm$, denoted by
$\tmmax=(\tmnmax,\tmxmax)$, can be calculated unambiguously for each
event.  These quantities are given by
\begin{eqnarray}\label{max1}
\tmnsqmax &=& \frac{\pya\cdot\pyb}{4\detM}\left[c_c-\frac{c_b^2}{c_a}\right]\,, \\ \label{max2}
\tmxsqmax &=& \frac{\pya\cdot\pyb}{4\detM}\left[c_c-\frac{(c_b+2\detM)^2}{c_a}\right]\,.
\end{eqnarray}
By construction, they are greater than the true masses.  Other
variables defined on the boundary can also be constructed.  For
example, if particular values of $\tmn$ are assumed, the extremal
values of $\tmx$, denoted $\tmx^{\rm min, max}(\tmn)$, can be obtained
using Eq.~(\ref{eq:quad}).  For $\tmn=\mn$, $\tmx^{\rm min(max)}(\mn)$
is smaller (larger) than $\mx$ by construction, with $\mx$ being the
upper (lower) endpoint of the distributions.  The relationship of
these quantities in a `typical' event is displayed in
Fig.~\ref{fig:typical}.  Note that $\tmx^{\rm min}(\mn)$ corresponds
to the quantity discussed in Ref.~\cite{Feng:1993sd}.  Since its
functional form is different from $\tmmax$, it contains in principle
complementary information.

Although not considered in this Letter, the methods for finding
  consistent $\tm$ and $\tmmax$ should be valid even when the equal
  mass constraints, Eq.~(\ref{eq:mass}), are relaxed.  In this case
  Eq.~(\ref{eq:quad}) becomes a quadratic function of two or three
  independent mass differences, for the case of one or no pairs of
  equal-mass particles, respectively.  A unique $\tmmax$, now
  containing three or four elements, can again be obtained
  analytically for each $\pmea$.

Note that $\tmmax$ depends only on $\pmea$, and so while the shape of
the distributions is sensitive to detailed dynamics and $\sqrt{s}$,
the position of the edges are not.  This should be compared with other
linear collider mass measurement techniques which depend on $\sqrt{s}$
being controllable/fixed, without including mass shell constraints.
For example, by varying $\sqrt{s}$, the threshold scanning method
\cite{AguilarSaavedra:2001rg} is sensitive to the production threshold
scale $2\mx$, while directly measuring $2\mn$ will be challenging
since $N/N'$ are invisible.  In addition, the distribution of
$E_{Y/Y'}$, the energy of $Y$ and $Y'$, have endpoints
\cite{Feng:1993sd, hep-ph/9910416}
\begin{equation}\label{eq:Eminmax}
  E^{{\rm max, min}}_{Y/Y'} =
  \frac{\sqrt{s}}{4}\left[1-\frac{m_{N}^2}{m_{X}^2}\right]\left[1\pm\sqrt{1-\frac{4m_{X}^2}{s}}\right]
\end{equation}
when radiation and detector smearing effects are neglected.  The true
mass $\truem$ can then be obtained if the endpoints and $\sqrt{s}$ are
accurately determined.  Depending on the values of $\truem$ and
$\sqrt{s}$, our method could have statistical advantages in the
endpoint determination.  Furthermore, the fact that the $\tmmax$ are
bounded from below by $\truem$ implies that these variables could be
particularly effective in separating the signal events from (the SM)
background when used simultaneously.  More interestingly, the
$\sqrt{s}$ independence and Lorentz invariance of $\tmmax$ leads to
the possibility of utilising these variables in hadron-hadron
collisions at the LHC, where the partonic $\sqrt{s}$ cannot be
controlled directly.  We shall illustrate these points with the
examples below.

\begin{figure}[!t]
  \begin{center}
    \scalebox{\scale}{
      \includegraphics{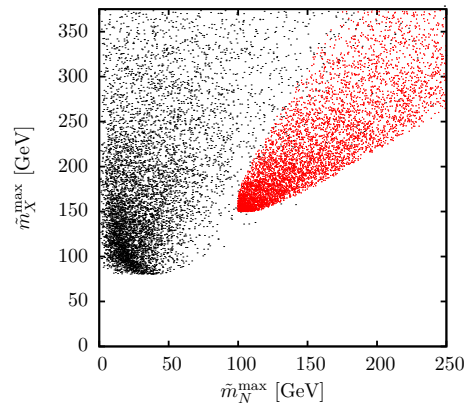}
    }
    \caption{Scatter plot for $\tmmax=(\tmnmax,\tmxmax)$ for the SM
      leptonic $\Wp\Wm$ events (black), and pair production of
      selectrons ($\mx=150$ GeV) in MSSM, followed by decays into
      neutralinos ($\mn=100$ GeV) and electrons (red) at a 3 TeV
      $e^+e^-$ collider.  10,000 events for each process are
      displayed.  No cuts, detector smearing and radiation effects are
      included.}\label{fig:maxmass}
  \end{center}
\end{figure}

\sectionhead{Examples} Our first example is based on a $e^+e^-$
collider with $\sqrt{s}=3$ TeV, the proposed CLIC energy \cite{CLIC}.
We use \herwigv \cite{Bahr:2008pv} to simulate pair production of
right handed selectrons ($\seR$) in MSSM, followed by decay into a
pair of electrons and two lightest neutralino ($\schi$), assumed to be
the superpartner of the SM $U(1)_{Y}$ gauge boson, and which is stable
and escape detection:
\begin{eqnarray}\label{eq:slepproc}
  e^+e^- &\to& \seR^+\seR^- \to e^+e^-\schi\schi\,.
\end{eqnarray}
The mass of $\seR$ ($\mx$) and $\schi$ ($\mn$) are chosen to be 150
and 100 GeV respectively.  The small electron mass means that
$\yya,\yyb$ in Eq.~(\ref{eq:lambdas}) can be safely neglected, leading
to much simplified analytic expressions.  For comparison, the
irreducible SM $\Wp\Wm$ background:
\begin{eqnarray}
  e^+e^- &\to& \Wp\Wm \to e^+e^-\nu\bar{\nu}\,,
\end{eqnarray}
is also simulated.  The cross sections for the two processes are
displayed in Table~\ref{tab:xsec}.

\begin{table}[t!]
  \centering
  \begin{tabular}{c|c|c}
    \hline
      & SM & MSSM \\
    \hline
    ($\mn,\mx$) [GeV] & (0, 80.4) & (100, 150)\\
    \hline
    $\sigma_{\rm total}$ $[{\rm fb}]$ & 7 & 68\\
    \hline
  \end{tabular}
  \caption{Total cross sections for $e^+e^-+\pmiss$ events for the SM
    $\Wp\Wm$ and MSSM selectron pair production, followed by decays
    into electrons and neutralinos at a 3 TeV $e^+e^-$ collider.  The
    $W \to e \nu_{e}$ branching ratio is taken as 0.108.}
  \label{tab:xsec}
\end{table}

In Fig.~\ref{fig:maxmass}, we show a scatter plot of $\tmmax$ for the
MSSM (red) and SM (black) processes at parton level, \ie without
initial and final state photon radiation.  While the cross section for
the MSSM signal process is already an order of magnitude larger than
the SM background, it is instructive to see that the two processes are
cleanly separated before applying additional selection cuts.  Given
the simple mass dependence, we expect similar scatter plots to be also
useful in separating different NP processes.

\begin{figure}[t!]
  \begin{center}
    \scalebox{\multiplotscale}{
      \includegraphics{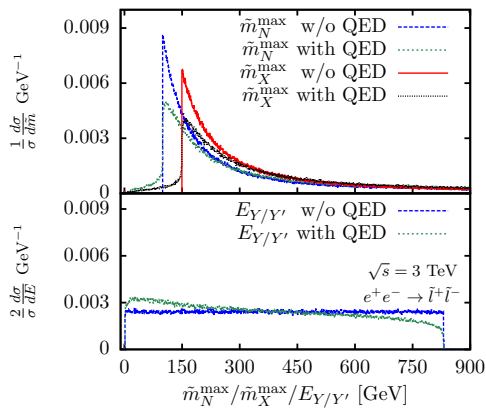} }
    \caption{$\tmnmax$ (blue/green), $\tmxmax$ (red/black)
      and $E_{Y/Y'}$ distributions for pair production of selectrons
      ($\mx=150$ GeV) in MSSM, followed by decays into electrons and
      neutralinos ($\mn=100$ GeV) at a 3 TeV $e^+e^-$ collider.
      Simulations both with and without the inclusion of QED radiation are
      displayed.  No cuts and detector smearing effects are
      included.}\label{fig:CLIC}
  \end{center}
  \vspace{-0.5cm}
\end{figure}

Next we present the $\tmnmax$, $\tmxmax$ and electron energy
$E_{Y/Y'}$ distributions for the MSSM sample in Fig.~\ref{fig:CLIC}.
In this particular sample, we see that at parton level there is a
statistical gain of a factor of 2 to 3 near the endpoint of $\tmnmax$
and $\tmxmax$ over that of the $E_{Y/Y'}$ distribution.  While the
large difference between $\sqrt{s}$ and $\tmmax$ results in the long
tails for the $\tmmax$ distributions, the tails fall off sufficiently
quickly and sharp edge structures remain at $\truem$.  Note that the
flat $E_{Y/Y'}$ distribution is due to the spin-0 nature of the
selectrons, and different spin assignments can lead to different
(endpoint) distributions.  Also, the small $4\mx^2/s$ ratio means that
$E^{\rm min}_{Y/Y'}$ is very close to zero, so this endpoint might not
be measured if additional energy/momentum cuts were imposed.

When bremsstrahlung effects are included, all distributions are
distorted.  Now $\pmiss$ cannot be fully determined, in part due to
initial state radiation down the beam pipe.  The $\pmiss$ values needed
for the more realistic distributions in Fig.~\ref{fig:CLIC} are
obtained from the momentum imbalance between the final state and the
initial state $e^+e^-$ systems, assuming {\it no} bremsstrahlung
effects for the latter.  Initial state radiation is calculable in
perturbative QED, and its effect on the parton $\tmmax$ distributions
may be incorporated in a more sophisticated treatment, which is beyond
the scope of the present study.  In our simple estimate, the $\tmmax$
distributions still display sharp edge structures around $\truem$
despite the radiation effects.

Next we turn to possible applications at the LHC.  For central
exclusive production (CEP) processes (see Ref.~\cite{Albrow:2010yb}
and references therein for more details), for example two-photon
production of a pair of charged particles ($X/X'$)
\begin{eqnarray}\label{eq:cep}
  pp &\to& p\, +\, \gam\gam\,+\,p \nonumber \\ && \phantom{p\,
    +\,\,\,\,} \gam\gam\to X^{(\pm)}X'^{(\mp)}\,,
\end{eqnarray}
followed by decays as depicted in Fig.~\ref{fig:proc}, all four
components of $\pmiss$ can be determined when the two final state
protons are measured, which could be achieved by installing proton
tagging detectors far from the interaction point~\cite{Albrow:2008pn}.
In the first equation of Eq.~(\ref{eq:cep}), the `$+$' signs represent
the presence of rapidity gaps.  Contrary to $e^+e^-$ processes,
$\sqrt{s}$ is differerent for each CEP event.  This means that the
$E_{Y/Y'}$ endpoint method cannot be directly used, while the $\tmmax$
method can.  The invariant mass/energy of $\gam\gam$ and $\pmiss$,
which have lower endpoints at $2\mx$ and $2\mn$ respectively, have
been proposed to measure $\truem$ in CEP \cite{Schul:2008sr}.  Since
$\tmmax$ takes the mass shell constraints into account, they are
expected to have sharper distributions over the other variables.  A
comparison between these observables, and the precision on $\truem$
that can be achieved at the LHC using $\tmmax$ will be discussed in a
separate article \cite{HarlandLang:2011ih}.

Finally, for inelastic processes at the LHC, only the transverse
components of $\pmiss$, \ie $\ptmiss$, might be measured.  If only the
short decay chain in Fig.~\ref{fig:proc} is observed, measuring
$\truem$ will be challenging.  In principle, $\truem$ could be
measured from the kink structure of $\mttwo^{\rm max}(\tmn)$
\cite{Lester:1999tx, arXiv:0706.2871, arXiv:0910.3679}.  However, the
kink resides at the tail of the $\mttwo(\tmn)$ distribution and so an
accurate measurement will be difficult.  In this case, the mass
measurement in CEP could be crucial.  It was shown in
Ref.~\cite{Cheng:2008hk} that $\mttwo(\tmn)$ is a boundary of the mass
region consistent with the mass shell constraints.  We have checked
numerically that this corresponds to $\tmx^{\rm min}(\tmn)$ over all
physical $\pmiss$ configurations, given $\ptmiss$.  How solutions
other than $\tmx^{\rm min}(\tmn)$ can be utilised (as discriminating
variables), and extending the methods presented to other event
topologies are subjects of on-going studies.

%%%%%%%%%%%%%%%%%%%%%%%%%%%%%%%%%%%%%%%%%%%%%%%%%%%%%%%%%%%%%%%%%%%
\begin{acknowledgments}
This work has been supported in part by the UK Science and Technology
Facilities Council.  WJS and LHL acknowledge support from an IPPP
Associateship.
\end{acknowledgments}

%%%%%%%%%%%%%%%%%%%%%%%%%%%%%%%%%%%%%%%%%%%%%%%%%%%%%%%%%%%%%%%%%%%

\end{document}